\documentclass[twocolumn,preprintnumbers,amsmath,amssymb]{revtex4-1}

\usepackage{color}
\usepackage{graphicx}
\usepackage{dcolumn}
\usepackage{amsthm}
\usepackage{bm}
\usepackage{siunitx}

\newcommand{\alphaeR}{0.89}
\newcommand{\alphaiR}{3.76}
\newcommand{\lzero}{\SI{280}{\micro\metre}}

\newcommand{\kB}{k_{\textup{B}}}
\newcommand{\me}{m_{\textup{e}}}
\newcommand{\vf}{v_{\textup{F}}}

\newcommand{\Lg}{L_{\textup{g}}}
\newcommand{\Lug}{L_{\textup{ug}}}
\newcommand{\LT}{L_{\textup{T1}}}
\newcommand{\Ls}{L_{\textup{strip}}}

\newcommand{\Te}{T_{\textup{e}}}
\newcommand{\TL}{T_{\textup{L}}}

\newcommand{\Vtg}{V_{\textup{tg}}}

\newcommand{\Vg}{V_{\textup{g}}}
\newcommand{\Va}{V_{\textup{a}}}
\newcommand{\Vb}{V_{\textup{b}}}
\newcommand{\Vth}{V_{\textup{th}}}

\newcommand{\Rh}{R_{\textup{h}}}
\newcommand{\Ih}{I_{\textup{h}}}

\newcommand{\taue}{\tau_{\textup{e}}}
\newcommand{\taui}{\tau_{\textup{i}}}
\newcommand{\alphae}{\alpha_{\textup{e}}}
\newcommand{\alphai}{\alpha_{\textup{i}}}

\newcommand{\p}{\textit{p}}
\newcommand{\np}{n_{\textit{p}}}
\newcommand{\Tp}{T_{\textit{p}}}
\newcommand{\Sp}{S_{\textit{p}}}

\newcommand{\baseT}{\SI{170}{\milli\kelvin}}

\newcommand{\mobility}{$\SI{172}{{\metre}^{2}\per{\volt\second}}$}
\newcommand{\density}{$n_0 = \SI{1.45e15}{\metre^{-2}}$}

\newcommand{\etal}{~\textit{et al.}~}
\newcommand{\Jurgen}{J{\"u}rgen K{\"o}nig}

\newcommand{\fig}[1]{Fig.~\ref{fig#1}}

\newcommand{\clfig}[1]{Figure~\ref{fig#1}}

\begin{document}


\title{Determining energy relaxation length scales in two-dimensional electron gases}


\author{Jordan Billiald$^1$, Dirk Backes $^1$, \Jurgen$^2$, Ian Farrer$^1$, David Ritchie$^1$, Vijay Narayan$^1$}%
\affiliation{%
$^1$Cavendish Laboratory, University of Cambridge, J. J. Thomson Avenue, Cambridge, CB3 0HE, United Kingdom. \\ $^2$Theoretische Physik and CENIDE, Universit{\"a}t Duisburg-Essen, 47048 Duisburg, Germany.}%

\date{\today}
\begin{abstract}
We present measurements of the energy relaxation length scale $\ell$ in two-dimensional electron gases (2DEGs). A temperature gradient is established in the 2DEG by means of a heating current, and then the elevated electron temperature $\Te$ is estimated by measuring the resultant thermovoltage signal across a pair of deferentially biased bar-gates. We adapt a model by Rojek and K\"{o}nig [Phys. Rev. B \textbf{90}, 115403 (2014)] to analyse the thermovoltage signal and as a result extract $\ell$, $\Te$, and the power-law exponent $\alphai$ for inelastic scattering events  in the 2DEG. We show that in high-mobility 2DEGs, $\ell$ can attain macroscopic values of several hundred microns, but decreases rapidly as the carrier density $n$ is decreased. Our work demonstrates a versatile low-temperature thermometry scheme, and the results provide important insights into heat transport mechanisms in low-dimensional systems and nanostructures. These insights will be vital for practical design considerations of future nanoelectronic circuits.
\end{abstract}

\maketitle

There currently exist well-established methods to probe the low-temperature (low-$T$) electrical and thermoelectric properties of two-dimensional electron gases (2DEGs). However, probing heat transport mechanisms in these systems has proven more challenging, primarily due to the lack of convenient low-$T$ thermometers that couple directly to the electron gas. Conventional low-$T$ thermometers such as germanium or ruthenium-oxide films are sensitive only to the lattice temperature $\TL$, the temperature of the crystal that hosts the 2DEG. At $\TL \lesssim \SI{1}{\kelvin}$ the coupling between electrons and phonons becomes relatively weak, and therefore $\Te$ can differ significantly from $\TL$. The electrical resistance of the 2DEG itself becomes insensitive to $\Te$ at these temperatures, since the majority of scattering events are from static impurities. These both therefore become ineffective at measuring $\Te$ in this regime. Accurate measures of $\Te$ can be obtained from the Coulomb-blockade characteristics of quantum dots~\cite{Davies_textbook} which are broadened at a finite $\Te$. The weak localization characteristics of 2DEGs~\cite{MittalSS1996} can also be useful as they depend sensitively on the phase-coherence length which is $\Te$-dependent. However, neither of these methods lend themselves easily to the measurement of spatial temperature gradients, which is required in order to measure the thermal conductivity $\kappa$~\cite{MolenkampPRL1992,SymeKellyPepper_JPhysC1989} or the energy relaxation length scale $\ell$ in 2DEGs.

Appleyard\etal\cite{AppleyardPRL1996} showed that the diffusive component of the thermopower $S$ can be used to detect differences between $\Te$ and $\TL$. Here $S \equiv \Vth/\Delta T$, where $\Vth$ is the thermovoltage developed in response to a temperature difference $\Delta T$. To do this they measured $\Vth$ across a pair of quantum point contacts (QPCs) with a heated electron gas between them, as was done similarly in Ref.~\cite{MolenkampPRL1990}, and more recently in Ref.~\cite{WiemannAPL2010}. Then $\Delta T$ was estimated as $\Vth/S$, where $S$ was obtained using the Mott relation~\cite{Mottformula}:

\begin{equation}
\label{Mottformula}
S = \frac{\pi^2\kB^{2}T}{3e\hbar^2}\left(\frac{\mbox{d}\ln\sigma}{\mbox{d}E}\right)_{E = \mu}.
\end{equation}

\begin{figure}
\centering
\includegraphics[width=2in]{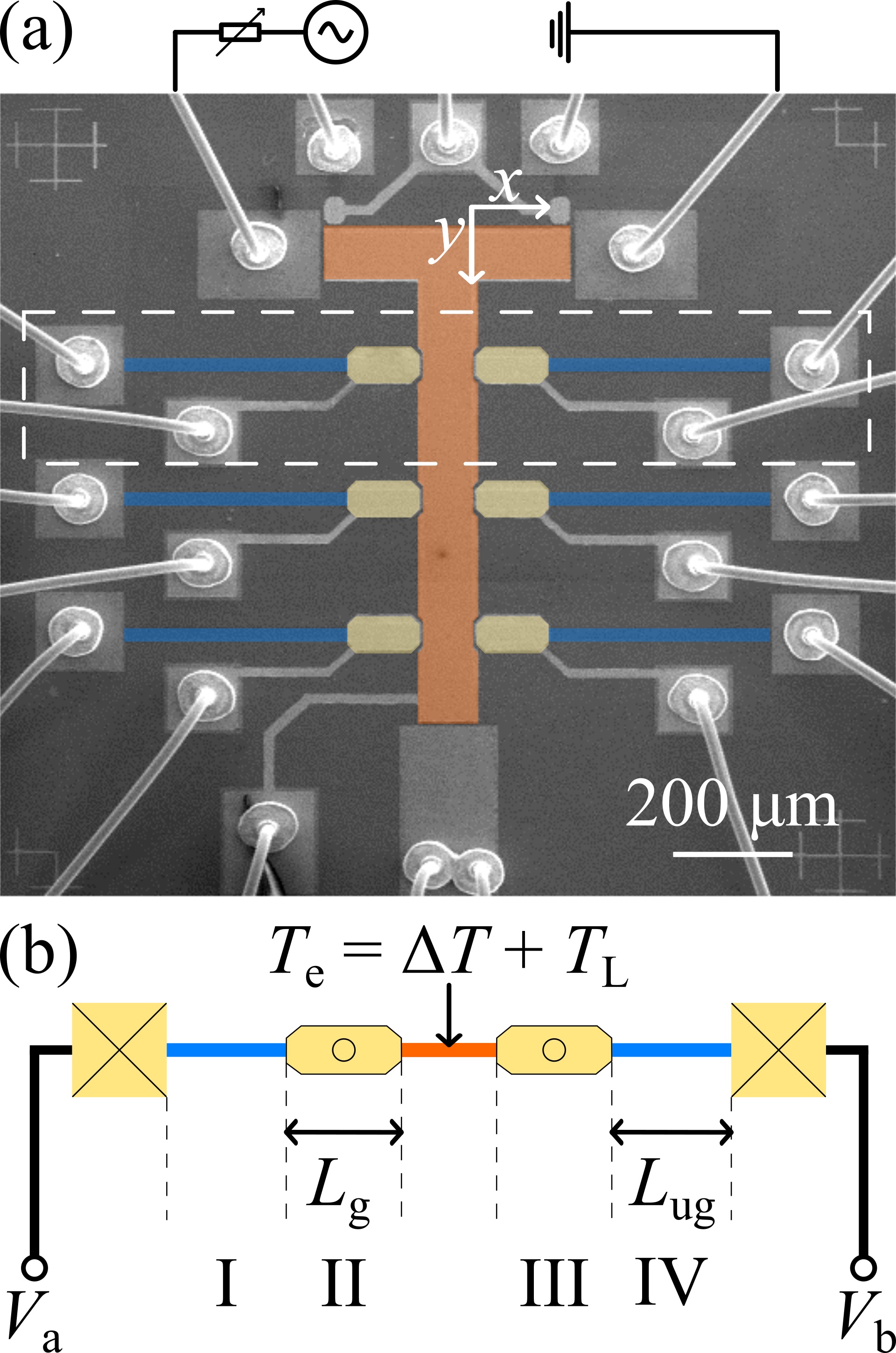}
\caption{Device geometry. (a) False-colour SEM image of the device used in our experiments. The first bar-gate thermometer (BGT) is outlined by white dashed lines. The three contacts at the top of the device allow a four-terminal measurement of the heating element resistance. (b) Schematic representation of a BGT which consists of gated (II and III, yellow) and ungated (I and IV, blue) regions of 2DEG, terminated by an ohmic contact. The orange section between the gates illustrates a region of `hot' electrons.}
\label{fig1}
\end{figure}

Here $\kB$ is the Boltzmann constant, $e$ is the electron charge, $\hbar \equiv h/2\pi$ with $h$ being Planck's constant, $\sigma$ is the electrical conductivity, $E$ is the total energy and $\mu$ is the chemical potential. This technique has been used to measure $\kappa$ in quantum wires~\cite{ChiattiPRL,MolenkampPRL1992}
and energy-loss rates in 2DEGs~\cite{ProskuryakovPRB2007}. Chickering\etal\cite{ChickeringPRL2009} showed Eq.~\ref{Mottformula} to be broadly valid in gated regions of a 2DEG between \SI{0.8}{\kelvin} and \SI{2}{\kelvin} and therein suggested the possibility of using a symmetric pair of bar-gates as a low-$T$ thermometer for the electron gas. This method was recently employed to measure $S$ in mesoscopic 2DEGs~\cite{NarayanPRB2012, NarayanJLTP2013, NarayanNJP2014}. Usefully, the relatively large size of the bar-gates eliminates the need for electron beam lithography which simplifies the fabrication process. However, it was noted in Ref.~\cite{ChickeringPRL2009} that the data systematically deviated from the Mott prediction. Rojek\etal\cite{RojekPRB2014} attributed these deviations to the spatial extent of the bar-gate thermometers (BGTs) being comparable to the energy relaxation length $\ell$ in the 2DEG and developed a model to account for this. In this work we adapt the model developed by Rojek and K\"{o}nig to refine the analysis of the signal produced by a BGT, and to make an accurate measurement of $\ell$.

\clfig{1}a shows a false-colour SEM image of a typical device. Please see the supplementary Material~\cite{SM} for wafer and fabrication details.
The device consists of a heating element and a longitudinal strip of 2DEG (of width \SI{100}{\micro\metre} and length $\Ls$~=~\SI{1}{\milli\metre}) which together form a `T' shape, with three BGTs along the strip. The first BGT is at a distance $\LT = \SI{200}{\micro\metre}$ from the heating element. A top-gate sits over the heating element and the strip and is used to tune the electron density $n$ in these regions. This design minimizes any power reflection at the interface between the heating element and the strip. The strip is terminated by a large ohmic contact.

Throughout the experiment all ohmic contacts through which no current is passed were assumed to be at $\TL$, since they are in direct thermal contact to the mixing chamber via the measurement wiring. \clfig{1}b shows a schematic of a single BGT. It is a symmetric structure consisting of two arms flanking the 2DEG strip. The left (right) arm is formed by a gated region labelled II (III), followed by an ungated region labelled I (IV), and terminated by an ohmic contact. The lengths of the gated and ungated regions of the arms are $\Lg = \SI{150}{\micro\metre}$ and $\Lug = \SI{455}{\micro\metre}$, respectively.

\begin{figure}
\centering
\includegraphics[width=3in]{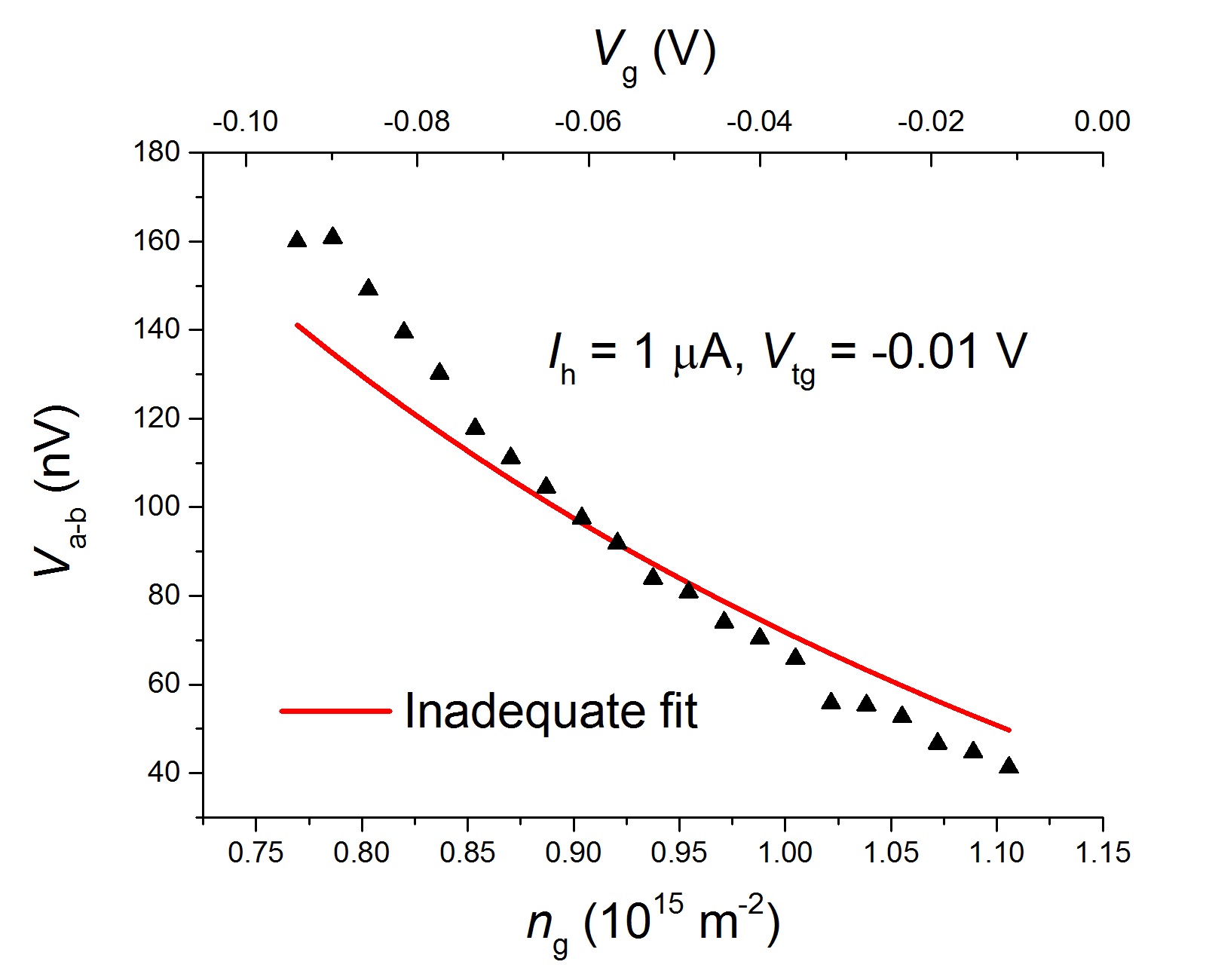}
\caption{The fit shown in this figure is based on a simple model where the hot electrons are assumed to relax to $\TL$ within the gated region. This model ignores any effects of the temperature profile along the BGT, and therefore the only fitting parameter is $\Te$. The fit is clearly inadequate with the fit systematically deviating from the experimental data. This kind of deviation was found to be typical across the parameter space of $\Ih$ and $\Vtg$ explored in this experiment.}
\label{fig2}
\end{figure}

The experiment involves passing a current $\Ih$ at frequency $f=\SI{10}{\hertz}$ through the heating element which Joule heats the electron gas with a power of $\Ih^2\Rh$, where $\Rh$ is the four-terminal electrical resistance of the heating element. This establishes a temperature gradient along the length of the 2DEG strip.
The thermovoltage $\Vth$ generated across a thermometer in response to $\Ih$ is detected at $2f$ using a lock-in amplifier. \clfig{2} shows $\Vth=\Va-\Vb$~(see \fig{1}b) from the first BGT while the gate voltage ($\Vg$) on gate II is swept. An almost identical result is obtained when gate III is swept, except that $\Vth$ is opposite in sign. The thermovoltages developed across the second and third BGTs were found to be negligible for even the largest used heating currents. Each section of the BGT from I to IV, contributes to $\Vth$:

\begin{equation}
\label{Thermovoltage}
\Vth = \sum_{\p}{\Sp \Delta \Tp}.
\end{equation}

Here $\p$ goes from I to IV, denoting each section shown in \fig{1}b, $\Delta \Tp$ is the temperature drop across section $\p$, and $\Sp$ for each section is given by the Mott formula for a non-interacting 2DEG:

\begin{equation}
\label{DrudeMottformula}
\Sp = \frac{\pi {\kB}^2 T_p m}{3e\hbar^2}\frac{(1 + \alphae)}{\np}.
\end{equation}

\begin{figure*}
\centering
\includegraphics[width=6.0in]{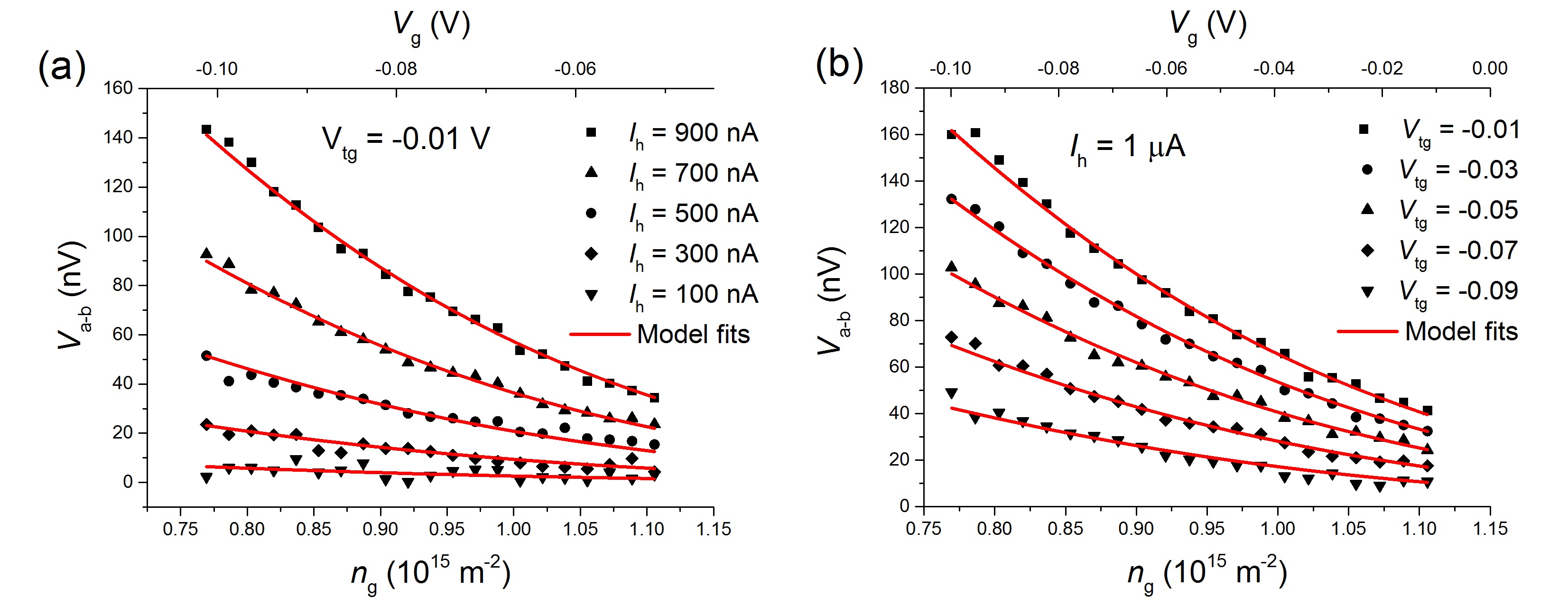}
\caption{Fits to the data using Eqs.~\ref{Thermovoltage},~\ref{DrudeMottformula} and \ref{Model}. The graphs show $\Vth$ as a function of the swept gate voltage $\Vg$ and $n_{\textup{g}}$, the corresponding carrier density under the swept gate. The different traces in (a) and (b) represent different heating currents $\Ih$ and different top-gate voltages $\Vtg$ respectively. The fits are markedly improved over, for example, that shown in \fig{2}.} 
\label{fig3}
\end{figure*}

\begin{figure*}
\centering
\includegraphics[width=6.0in]{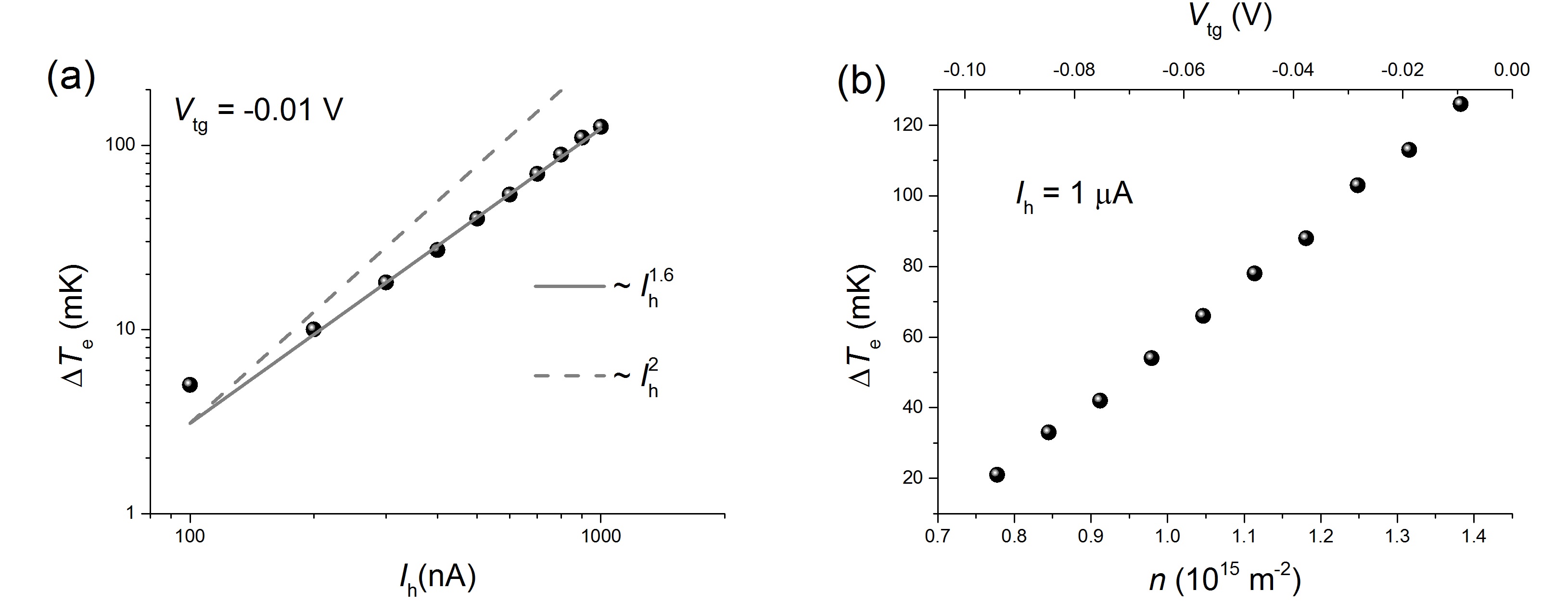}
\caption{(a) The elevated electron temperature $\Delta \Te \equiv \Te-\TL$ extracted from the model is seen to increase as a power-law of $\Ih$. The broken line shows an $\Ih^2$ trend, which would be the case if $\Delta \Te$ increased linearly as a function of applied heating power. The sub-linear dependence on power indicates that there are heat losses to the surrounding lattice. (b) As $\Vtg$ becomes more negative, $\Te$ at the first thermometer decreases rapidly, indicating that the electrons thermalize over a much shorter distance.}
\label{fig4}
\end{figure*}

Equation~\ref{DrudeMottformula} is obtained by substituting the Drude expression $\sigma = \np e^2 \taue/m$ for the electrical conductivity into Eq.~\ref{Mottformula}. Here $\np$ is the 2D number density of charge carriers in section $\p$, $\taue$ is the Drude elastic scattering time, $m$ is the effective mass of the charge carriers ($=0.067\me$ in GaAs-based 2DEGs, with $\me$ being the bare electron mass), $\alphae \equiv (\np/\taue)(\mbox{d}\taue/\mbox{d}\np)$, and $\Tp$ is the average electron temperature in region $p$.

\clfig{2} also shows the best fit to the data by assuming that hot electrons relax to $\TL$ within a distance $\Lg$ and therefore over the gated region, regardless of the carrier density ${n_{\textup{g}}}$ beneath the gate. In this simple picture the temperature profile along the BGT arm is not relevant, and the only unknown is $\Te$ between the BGT arms. However the quality of the fit is clearly inadequate and this is found to be the case across the parameter space of $\Ih$ and $\Vtg$ explored in this experiment.

We first describe why it is essential to consider the contribution from \textit{all} the regions $p =$~I to IV towards $\Vth$. At low-$T$ and especially in high-mobility 2DEGs, the energy relaxation length $\ell$ over which hot electrons relax through inelastic processes can greatly exceed the mean-free path of electrons~\cite{GanczarczykPRB2012,RojekPRB2014}. The dependence of $\ell$ on the inelastic scattering time $\taui$ is given by $\ell~\equiv~\sqrt{D \taui}$, where $D={\vf}^2 \taue/2$ is the diffusion constant of the electrons, and $\vf$ is the Fermi velocity. $\taui$ has a power-law dependence on $n$: $\taui = \tau_{i,0}(n/n_0)^{\alphai}$, where the subscript $0$ denotes the values in the ungated 2DEG. Thus, the distance over which electrons lose their excess energy in the BGT arm is crucially dependent on $n_p$. The ohmic contact enforces $\Te = \TL$ at the 2DEG to ohmic contact interface, and this needs to be taken into consideration if $\ell$ is comparable to $\Lg+\Lug$. Therefore in all the above situations, both the gated and ungated regions contribute to $\Vth$ in a manner dependent on $\Te$ at the centre of the BGT, $\Te$ at the interface of the gated and ungated regions, $\ell$, and $\alphai$, all of which need to be estimated self-consistently. To do this we adapt the model used in Ref.~\cite{RojekPRB2014} (described in the Supplementary Material~\cite{SM}). The resulting expression for $\Delta \Te \equiv \Te - \TL$ at the junction of the gated and ungated regions $\Delta \Te(\Lg)$, reads:
\begin{widetext}
\begin{equation}
\label{Model}
\Delta \Te(\Lg)= \Delta \Te \frac{z \sinh(\Lug/\ell_0)}{\cosh(\Lug/\ell_0)\sinh(\Lg/\ell) + z\sinh(\Lug/\ell_0)\cosh(\Lg/\ell)}.
\end{equation}
\end{widetext}

Here $\ell_0$ is the energy relaxation length at $n=n_0$, and $z \equiv (n/n_0)^{(1+\alphae-\alphai)/2}$.
Within the framework of Rojek\etal's linearized model~\cite{RojekPRB2014}, $\Tp = \TL$. Equation~\ref{Model} provides an expression for $\Delta \Tp$, which when substituted together with Eq.~\ref{DrudeMottformula} in Eq.~\ref{Thermovoltage}, results in an expression for $\Vth$ as a function of $n_0$, $n$ in each gated region, $\ell_0$, $\alphae$, $\alphai$ and $\Te$. We measure $n_0$ and $n(\Vg)$ in the device by observing $\Vg$-dependent edge-state reflections in the quantum Hall regime~\cite{BaenningerPRBR2005}, and $\alphae$ is extracted from the dependence of $\sigma$ on $n$ and turns out to be $\approx\alphaeR$ over the relevant range of $n$. This leaves three unknowns, namely $\ell_0$, $\Te$, and $\alphai$, which are used as fitting parameters. Importantly though, by fitting to several complementary data sets whilst varying different experimental parameters such as $\Ih$ and $\Vtg$, we are able to considerably reduce the uncertainty in these three fitting parameters (see Supplementary Material~\cite{SM}).

\clfig{3}a shows $\Vth$ against ${n_{\textup{g}}}$ for varying $\Ih$ which will produce different $\Delta \Te$. \clfig{3}b shows $\Vth$ against ${n_{\textup{g}}}$ but for varying $\Vtg$ which will vary $\ell$ in the 2DEG strip as well as $\Delta \Te$ via its effect on $\Rh$. Clearly, the model produces excellent fits to the data with no discernible systematic deviations. Similar data and quality of fit is obtained when the adjacent BGT is swept.

The results of the fitting are the $\Delta \Te$ for each $\Ih$ and $\Vtg$ in the two data sets. \clfig{4}a shows $\Delta \Te$ as a function of $\Ih$ on a log-log scale, and we find that $\Delta \Te \propto \Ih^{1.65}$. This sub-squared dependence is presumably due to a fraction of the power being lost to the lattice and to the ohmic contacts. The values of $\alphai$ and $\ell_0$ are found to be $\approx\alphaiR$ and $\approx\lzero$, respectively. \clfig{4}b shows $\Delta \Te$ as a function of $\Vtg$ after $\Delta \Te$ has been scaled for the changing value of $\Rh$. This has been done by applying a corrective factor of $R_{\textup{h},0}/R_{\textup{h}}(\Vtg)$ so that any change in $\Delta \Te$ is now due solely to a change in $\ell$. Here $R_{\textup{h}}(\Vtg)$ is the electrical resistance of the heating element as a function of $\Vtg$. Therefore the graph suggests that for $\Vtg \lesssim \SI{-0.11}{\volt}$
, the hot electrons completely relax within a distance $\LT$. 

The data therefore suggests that hot electrons thermalize over macroscopic length scales ($\approx \SI{300}{\micro\metre}$) in the ungated 2DEG, but that this length scale rapidly decreases with $n$. This strongly justifies the need to account for the ungated 2DEG arms when using BGTs at high $n_{\textup{g}}$. This is also consistent with the negligible thermovoltage detected across the second and third BGTs which are at distances of $\SI{500}{\micro\metre}$ and $\SI{800}{\micro\metre}$ from the heating element, respectively. While $n_{\textup{g}}$ can certainly be lowered untill $\ell < \Lg$ such that $\Delta \Te(\Lg)=0$, care must be taken to ensure that Eq.~\ref{DrudeMottformula} remains valid. Indeed, we have observed that when the 2DEG approaches the localized regime (when $\sigma$ becomes $\lesssim 3e^2/h$) the model is unable to fit the data satisfactorily. In this experiment $n$ was conservatively limited to $\geq \SI{0.7e15}{\metre^{-2}}$ corresponding to $k_{\textup{F}} l \geq 4$ and $r_{\textup{s}} \leq 2$, where $k_{\textup{F}}$ is the Fermi wave vector, $l$ is the elastic mean free path, and $r_{\textup{s}}$ is the interaction parameter defined as the ratio of the Coulomb energy and kinetic energy of the 2DEG.

\begin{figure}
\centering
\includegraphics[width=3in]{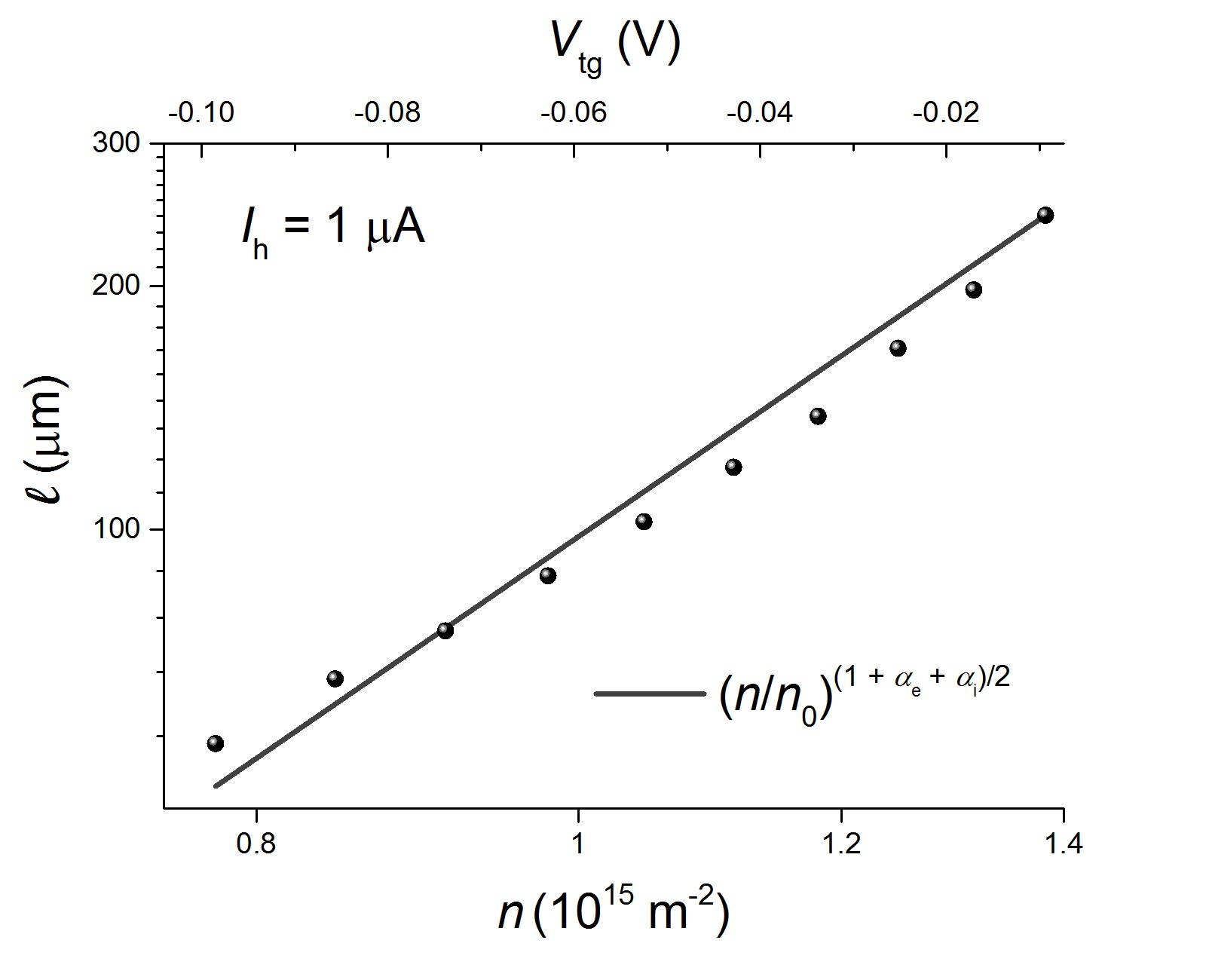}
\caption{Energy relaxation length $\ell$ as a function of carrier density $n$. The data in \fig{4}b produces an independent measure of $\ell(n)$ which we find to be in excellent agreement with the model exponents.}
\label{fig5}
\end{figure}

Importantly, the model \textit{only} provides the value of $\ell_0$ and $\alphai$ from which $\ell(n)$ can be reconstructed. However, as argued in the previous paragraphs, this information is independently contained in \fig{4}b. The $\Te$-profile in the 2DEG strip is given by (see Supplementary Material~\cite{SM}):

\begin{equation}
\label{TevsVg}
\Delta \Te(y) = \Delta T_{\textup{e,0}} \frac{\sinh{[(\Ls - y) / \ell]}}{\sinh{(\Ls /\ell)}}
\end{equation}

Here $y$ is the spatial coordinate along the 2DEG strip (see \fig{1}b) and $\Delta T_{\textup{e,0}} \equiv \Delta \Te(y=0)$, the temperature elevation in the heating channel. Since the heating channel is effectively at a constant temperature between pairs of points in \fig{4}b, the ratio $\Delta \Te(n_1)/\Delta \Te(n_2)$ taken at $y=\LT$ provides an implicit relation between $\ell(n_1)$ and $\ell(n_2)$, where $n_1$ and $n_2$ are the respective carrier densities. Therefore this allows the inference of $\ell$ from $\Te$, and reconstruction of $\ell$ as a function of $n$ as shown in \fig{5}. This is the main result of this study showing the dependence of the energy relaxation length scale on the carrier density. We stress that this is a \textit{direct} measurement of $\ell(n)$ using the pre-calibrated BGT, which we find to be in striking agreement with indirectly obtained dependence $\ell(n) = \ell_0(n/n_\textup{0})^{(1 + \alphae + \alphai)/2}$, using the values of $\ell_0$ and $\alphai$ derived from the model and the fitting.

To summarize, we have demonstrated that BGTs are a versatile tool with which to detect elevated electron temperatures. As remarked earlier, BGTs are an attractive alternative to QPCs. QPC fabrication requires electron beam lithography and they come with the associated difficulties of sub-micrometre devices such as sensitivity to electrical shock and vulnerability to disorder. In contrast, BGTs can be macroscopic and therefore free from the above mentioned difficulties.

While our manuscript seems to suggest that the trade-off between QPCs and BGTs is that detailed modelling is required to extract $\Te$, it is important to note that modifying the device design such that $\Lg > \ell$ significantly reduces the complexity of the analysis. On the other hand, the advantages of the employed device design are that it allows for the determination of $\ell(n)$ using a single BGT. We note that using a second BGT further displaced along the length of the 2DEG strip should also allow for such a measurement, so long as $\ell(n)<L_\textup{T2}$.

We acknowledge funding from the Leverhulme Trust, UK and the Engineering and Physical Sciences Research Council (EPSRC), UK. JB and VN acknowledge helpful discussions with Chris Ford and Charles Smith. Supporting data for this paper is available at the DSpace@Cambridge data repository (https://www.repository.cam.ac.uk/handle/1810/248776).


\setcounter{figure}{0} \renewcommand{\figurename}{Fig. S}

\onecolumngrid

\vspace{12pt}

\begin{center}

\large \textbf{Determining energy relaxation length scales in two-dimensional electron gases \\ Supplementary Material}

\end{center}

\twocolumngrid

\section{Wafer, device fabrication, and measurement system details}

The 2DEGs used in this study have a mobility of \mobility~at a carrier density of \density. For a top-gate voltage of $\Vtg=\SI{0}{\volt}$, $\Rh=\SI{230.9}{\ohm}$ which corresponds to a resistivity of $\SI{44.4}{\ohm}$/Sq. The typical ohmic contact resistance is $\approx\SI{200}{\ohm}$ as determined by the difference between two and four-terminal resistances (corrected for known cryostat wiring resistances). Therefore the ohmic contacts should not effect thermalization.
A wet etch is used to define the conductive mesa, and Au-Ge-Ni ohmic contacts and Ti-Au top-gates are then deposited by thermal evaporation. Photolithography is used at each stage to define the layer pattern in photoresist on the substrate.
Measurements on the devices were performed in a dilution refrigerator with a base temperature of \baseT. A ruthenium oxide thermometer attached to the mixing chamber of the dilution refrigerator was used to measure $\TL$.

\section{Thermopower diffusion model}

\begin{figure*}[t]
\centering
\includegraphics[width=6.5in]{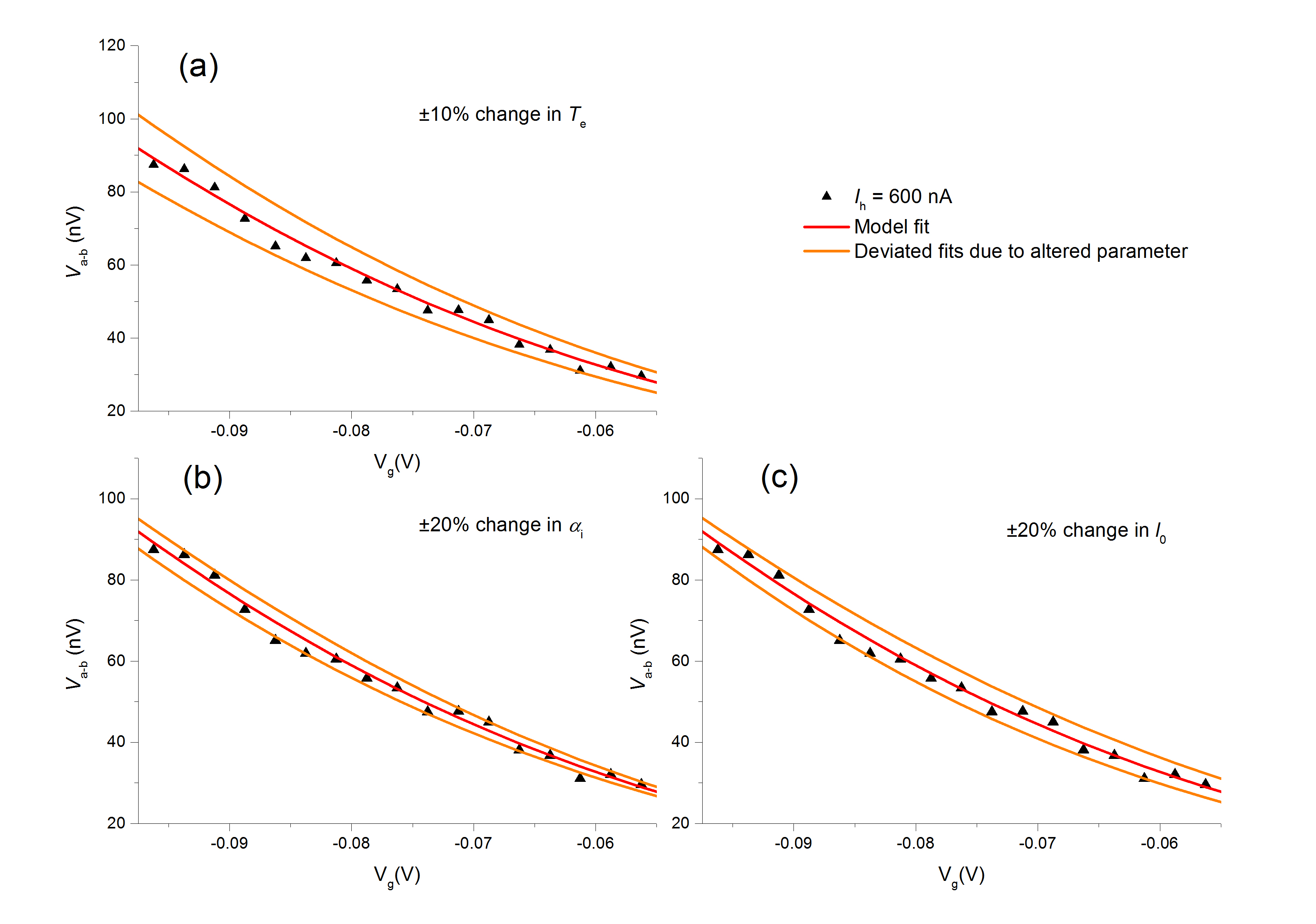}
\caption{Fits to a set of data showing the sensitivity of the fit to altered fitting parameters. }
\label{SM_fig1}
\end{figure*}

Rojek~\textit{et al.} consider a fixed electrical power input into a two-dimensional electron gas (2DEG) that results in an elevated electron temperature $\Te\equiv \Delta \Te+\TL$ with respect to the lattice temperature $\TL$. In the steady-state, the power input to the 2DEG is balanced by the heat loss to the lattice. Then, one can derive from the continuity equation for heat and charge flow a diffusion equation for $\Delta \Te (x)$,

\begin{equation} \tag{S1}
\Delta \Te - \frac{\ell^2}{\kappa}\frac{\partial}{\partial x}\left( \kappa \frac{\partial \Delta \Te}{\partial x} \right) = 0.
\label{DiffusionEquation}
\end{equation}

Here $\kappa$ is the thermal conductivity of the 2DEG, $\ell$ the energy relaxation length, and $x$ is the spatial coordinate along which the temperature varies (for the temperature variation within the BGT, relevant for Eq.~4, this is the $x$-direction while for the temperature variation within the 2DEG strip, relevant for Eq.~5,  this is the $y$-direction, see Fig.~1 of the main text).

An applied gate voltage changes the electron density $n$ which, in turn, modifies the transport properties, in particular the elastic and inelastic scattering times $\taue$ and $\taui$, the thermal conductivity $\kappa$, and the energy relaxation length $\ell$.
Their $n$-dependences are described by power laws with the following exponents:

\begin{align}
&\mbox{d}\ln\taue / \mbox{d} \ln n=\alphae, \nonumber \\
&\mbox{d}\ln\taui / \mbox{d} \ln n=\alphai, \nonumber \\
&\mbox{d}\ln\kappa / \mbox{d} \ln n=1+\alphae, \nonumber \\
&\mbox{d}\ln\ell / \mbox{d} \ln n=(1+\alphae+\alphai)/2. \nonumber
\end{align}

For each region with constant electron density (the gated region II(III) and the ungated region I(IV) of the BGT, relevant for Eq.~4, as well as the 2DEG strip, relevant for Eq.~5), the general solution of Eq.~\ref{DiffusionEquation} is
\begin{equation} \tag{S2}
\label{TProfile}
\Delta \Te(x) = a e^{-x/\ell} + b e^{x/\ell}  \, ,
\end{equation}
where the coefficients $a$ and $b$ have to be determined by boundary conditions.

\subsection{Derivation of Eq.~4}

For the BGT, the spatial position $x$ is defined relative to the interface between the 2DEG strip and the gated region II. 
There are two regions of constant electron density (the gated and the ungated one, I and II) which requires four boundary conditions to fully determine the temperature profile.
One is given by $\Delta \Te(0)$ at $x=0$, the quantity to be measured by the BGT.
Furthermore, the Ohmic contact at $x=L_{\textup{g}} + L_{\textup{ug}}$ fixes the electron temperature to the lattice temperature, $\Delta \Te(L_{\textup{g}} + L_{\textup{ug}})=0$.
Finally, at $x=L_{\textup{g}}$ (the interface between the gated (I) and the ungated (II) region), both the temperature $\Delta \Te$ and the heat current $-\kappa(\partial \Delta \Te/\partial x)$ have to be continuous.
It is, then, a matter of solving a set of four linear equations to express the coefficients of Eq.~\ref{TProfile} in terms of $\Delta \Te(0)$.
However, the only relevant feature of the full temperature profile that enters the thermovoltage is the temperature at the interface between region I and II, $\Delta \Te(\Lg)$.
Plugging $x=\Lg$ into the full solution immediately leads to the compact formula Eq.~4, where $z \equiv (\kappa/\ell)/(\kappa_0/\ell_0) = (n/n_0)^{(1+\alphae-\alphai)/2}$ originates from the boundary condition of the heat current to be continuous.

In the limit of a large gated region, $\Lg \gg \ell$, the coefficient $b$ of the exponentially increasing term in Eq.~\ref{TProfile} vanishes, $\Delta \Te(\Lg)$ goes to zero, and the ungated region does not influence the measure thermovoltage.
For $\Lg \lesssim \ell$, however, the ungated region becomes important. 
Ignoring it leads to an inadequate fit to the data, as shown in Fig.~2 of the main text.

\subsection{Derivation of Eq.~5}

The modelling of the temperature profile along the $y$-direction in the 2DEG strip is analogous to that of the BGT. Since the electron density is constant in the 2DEG strip, only two boundary conditions are needed.
They are given by fixing $\Delta \Te$ to $\Delta T_{\textup{e,0}}$ at $y=0$ and to $0$ at the Ohmic contact $y = L_{\textup{strip}}$.
Formally, determining the temperature profile along in the 2DEG strip is just a special case of the calculation for the BGT.
Therefore Eq.~5 follows from Eq.~4 by performing the replacements $z \rightarrow 1$, $\ell_0\rightarrow \ell$, $\Lg \rightarrow y$, $\Lug \rightarrow L_{\textup{strip}} - y$, and $\Delta \Te \rightarrow \Delta T_{\textup{e,0}}$.

\section{Sensitivity of fit on parameters}

Figure S 1 shows the sensitivity of a model fit to the fitting parameters. The particular data used to demonstrate this corresponds to $\Ih=\SI{600}{\nano\ampere}$ at $\Vtg=\SI{0}{\volt}$, although similar results are obtained for all other values of $\Ih$ and $\Vtg$ used in this experiment. The optimum fit is contrasted to the resulting fit when a specific fitting parameter is changed by a given percentage. In panel (a) $\Delta \Te$ was varied by 10\% in each direction, in panel (b) and (c), $\alphai$ and $\ell$ were varied by 20\% in each direction respectively. The result is a clearly deviated and inadequate fit in each case. However these percentage changes are chosen such that the quality of the fit visibly deteriorates. In fact the residual of the fit is more sensitive to changes in a single fitting parameter, for example, doubling for a 5\% change in $\Delta \Te$. This highlights the technique's usefulness as an electron thermometer.

\end{document}